\documentclass[twocolumn,twoside,preprintnumbers,amsmath,amssymb,pacs]{revtex4}
\usepackage{epsfig}
\usepackage{graphicx}
\usepackage{dcolumn}
\usepackage{bm}
\usepackage{amssymb}

\usepackage{times}
\usepackage{epsf}

\usepackage{axodraw}

\usepackage{fancyhdr}

\usepackage{pslatex}

\newcommand{\en}{\end{equation}}
\newcommand{\ena}{\end{eqnarray}}
\newcommand{\be}{\begin{equation}}
\newcommand{\ba}{\begin{eqnarray}}
\newcommand{\ea}{\end{eqnarray}}
\newcommand{\beqa}{\begin{eqnarray}}
\newcommand{\eeqa}{\end{eqnarray}}
\newcommand{\beqn}{\begin{equation}}
\newcommand{\eeqn}{\end{equation}}

\newcommand{\<}{\langle}
\renewcommand{\>}{\rangle}

\newcommand{\beq}{\begin{equation}}
\newcommand{\eeq}{\end{equation}}

\def\F2{\langle \alpha_s F_{\mu\nu} F^{\mu\nu} \rangle}

\def\A2{\langle A_{\mu}^a A_{\mu}^a \rangle}

\def\Z1{\widetilde{Z}_1}

\def\be{\begin{equation}}
\def\ee{\end{equation}}
\def\bea{\begin{eqnarray}}
\def\eea{\end{eqnarray}}
\def\A{{\mathcal A}}

\def\LZ1{\widetilde{{\mathcal Z}}_1}
\def\<{\langle}
\def\>{\rangle}

\newcommand{\hbo}{\hbox to 1 true cm {\hfill } }
  
\begin{document}

\title{{\Large Propagators, Running Coupling and Condensates in Lattice QCD}}

\author{Attilio Cucchieri and Tereza Mendes}

\affiliation{Instituto de F\'\i sica de S\~ao Carlos, Universidade de S\~ao Paulo,
                \\ C.P.\ 369, 13560-970, S\~ao Carlos, SP, Brazil }

\received{on 24 March, 2006}

\begin{abstract}

We present a review of our numerical studies of the running coupling constant,
gluon and ghost propagators, ghost-gluon vertex and ghost condensate
for the case of pure $SU(2)$ lattice gauge theory in the
minimal Landau gauge. Emphasis is given to the infrared regime, in order
to investigate the confinement mechanisms of QCD. We compare our
results to other theoretical and phenomenological studies.



\end{abstract}

\maketitle

\thispagestyle{fancy}
\setcounter{page}{0}


\section{Introduction}

The strong force --- one of the four fundamental interactions
of nature along with gravity, electromagnetism and the
weak force --- is the force that holds together protons
and neutrons in the nucleus. The strong interaction is
described by Quantum Chromodynamics (QCD) \cite{moriyasu}.
This description is based on a model of elementary particles ---
the quarks --- possessing ``color charge'' and interacting
through the exchange of gauge fields --- the gluons
(equivalent to the photons in the electromagnetic interaction).
QCD is a quantum field theory, with local $SU(3)$ gauge symmetry,
corresponding to three possible colors.
The fact that the gauge group of QCD is non-Abelian implies that the
gluons possess color charge and therefore interact with each other,
as opposed to the photons.
The only parameters of the theory
are the masses of the various types (called ``flavors'') of
quarks considered and the value of the strong coupling constant.

A unique feature of the strong force is that the particles that feel it
directly --- quarks and gluons --- are completely hidden from us,
i.e.\ they are never observed as free particles.
This phenomenon, known as confinement, makes
QCD much harder to handle theoretically than the theories
describing the weak and electromagnetic forces.
Indeed, the coupling constant $\alpha_s$ of the strong
interaction becomes negligible only in the limit
of small distances, or equivalently in the limit of high energy or
momentum. This property is called asymptotic freedom.
At larger distances (i.e.\ smaller energies) there is an increase
in the intensity of the interaction and it is believed that
the force of attraction between quarks is constant, for sufficiently
large distances, determining the confinement of quarks and gluons inside the
hadrons.
The fact that $\alpha_s$ is not negligible at low energies makes the
study of important phenomena such as the mechanism of quark and gluon confinement,
the hadron mass spectrum and the deconfining transition at finite
temperature inaccessible to calculations using perturbation theory.
These phenomena must therefore be studied in a nonperturbative
way.

The nonperturbative study of QCD is possible in the lattice
formulation of the theory, introduced by Wilson in 1974 \cite{lattice}.
This formulation offers a convenient nonperturbative
regularization, preserving the theory's gauge invariance.
The essential ingredients for the lattice formulation are:
1) path-integral quantization, 2) continuation to imaginary
or Euclidean time and 3) lattice regularization (given by the discretization
of space-time).
The combination of the first two ingredients makes the theory equivalent to
a model in classical statistical mechanics: indeed,
in Euclidean space a path integral for the quantum theory is equivalent to
a thermodynamic average for the corresponding statistical mechanical system.
For QCD, the square of the bare coupling constant $g_0$ of the field theory
corresponds directly to the temperature $1/\beta$ of the statistical
mechanical model.

The third ingredient --- the lattice discretization --- represents an
ultraviolet regularization.
In fact, the lattice spacing $a$ corresponds to a high-momentum cutoff,
since momenta higher than $\sim 1/a$ cannot be represented on the lattice.
In this way the ultraviolet divergences, appearing in the calculation of
physical quantities, are suppressed and the theory is
well defined. Of course, in order to recover the continuum-space theory we must take
the limit $a \to 0$.
In this process it is necessary to ``tune'' the bare parameters of the theory
in such a way that
physical quantities converge to
finite values, which can then be compared to experiment.
In particular, in the limit $a \to 0$, a correlation length $\xi$
measured in units of the lattice spacing, i.e.\ $\xi/a$, must go to infinity.
In other words, the lattice theory considered must approach a critical
point, i.e.\ a second-order phase transition.
Thus, the study of the continuum limit in quantum field theories on the
lattice is analogous to the study of critical phenomena in statistical
mechanics.
The correspondence between Euclidean field theories and classical statistical
mechanics allows the application of usual statistical-physics methods
to the study of QCD.
In particular, one may perform numerical simulations by Monte Carlo
methods, which are based on a stochastic description of the systems
considered \cite{MC}.

Despite the similarity of the methods, the Monte Carlo simulation of gauge theories
is much more complex than in the case of the usual statistical mechanical models,
requiring great computational effort and specific numerical techniques for the
production of the data. Moreover,
we must consider three limits in order to obtain the desired physical results from
the simulation data: 1) the infinite-volume limit (or thermodynamic limit),
2) the continuum limit (i.e.\ the value of $a$ must be sufficiently small when compared to the
relevant distance for the problem) and 3) the chiral limit (in order to consider
physical values for the masses of the light quarks).
The above limits are not independent, since to get to the continuum limit
and to be able to consider small masses for the quarks one needs a 
sufficiently large number of lattice points (corresponding to a small enough
lattice spacing and to a large enough physical size of the lattice),
which increases considerably the computational effort.

The study of lattice QCD constitutes a so-called {\em Grand Challenge} computational
problem \cite{culler}. Indeed,
simulations of full QCD --- i.e.\ including effects
of dynamical fermions --- for quark masses in the region of physical
values are still extremely slow and they are in general carried out on
supercomputers, involving the effort
of large collaborations such as the UKQCD in the United Kingdom and
the JLQCD in Japan.
Also, several research groups have built
QCD-dedicated computers, using parallel architecture.
Examples are the {\tt Hitachi/CP-PACS} machine at the
University of Tsukuba in Japan \cite{kanaya:2001aq},
the {\tt QCDSP} and {\tt QCDOC} machines at
Columbia University in the USA \cite{Columbia},
and the {\tt APE} machines \cite{APE} at
various research centers in Europe. These computers
range from about 1 to 10 {\tt teraFLOPS}. In addition to these
large projects, many groups base their simulations on
clusters of workstations or personal computers (PC's)
\cite{Luscher:2001tx}.
These systems do not yet provide the same efficiency in parallelization
as the machines with parallel architecture, but their cost is much
lower.
In addition to the computational power, the numerical and analytical 
techniques used in the simulations and in the interpretation of
the produced data are of great importance in the field.
Significant progress has been achieved through the development of
more efficient simulation algorithms, new methods for interpolation and
extrapolation of the numerical data and a better understanding of
the systematic effects to which the simulation may be subject, such
as finite-volume effects and discretization errors.
Progresses in the field are reported annually at the {\em Lattice} 
conference \cite{latticeproc}.

Despite the great computational difficulty, numerical studies
of QCD have already provided important contributions to the study
of the strong force \cite{Creutz:2003qy}.
In particular, numerical simulations of QCD
are now able to produce calculations of the strong
coupling constant $\alpha_s(\mu_0)$, taken at a fixed
reference scale $\mu_0$, with precision comparable
to the experimental one or better \cite{attilio}. These results are presently included
in the world average for this quantity \cite{Hinc}.
Also, the mass spectrum of the light hadrons (including the two light quarks and the
strange quark) has been determined (with great precision) \cite{cppacs}
for the quenched case, in which
the configurations are produced without considering effects of dynamical quarks.
One does not obtain complete agreement with the experimental spectrum, but the
observed discrepancies are of at most 10\%.
Similar calculations are now being performed for the full-QCD case.
Finally, lattice simulations constitute the only known evidence
for the quark-deconfining transition at finite temperature \cite{karsch}
and its predictions are of direct interest for the current experiments
in search of new states of matter in the laboratories Brookhaven
and CERN.

There is presently great interest in the results of the simulations
described above and one hopes to be able to solve many theoretical
questions about QCD and the standard model \cite{wilczek}.
Indeed, lattice-QCD simulations
are now able to provide quantitative predictions with errors of a
few percent. This means that these simulations will soon
become the main source of theoretical results for comparison
with experiments in high-energy physics \cite{Hprecision}, enabling a much
more complete understanding of the physics of the strong force.


\subsection{Lattice QCD at the IFSC--USP}

Since the beginning of 2001 we have been carrying out a project
on numerical simulations of lattice gauge theories at the 
Physics Department of the University of S\~ao Paulo in S\~ao Carlos 
(IFSC--USP), funded by FAPESP \cite{noi}.
The project included the installation of 2 PC clusters (with a total of 28
processing nodes). The resulting computer power is of approximately
$40$ {\tt gigaFLOPS} for peak performance.
We have performed production runs since July of 2001 and have started
intensive parallel simulations in November 2002.
Our main research topic (see Sections \ref{sec:conf}--\ref{sec:runn} below)
is the investigation of the infrared
behavior of various propagators and vertices in Landau gauge with the goal of
verifying the so-called Gribov-Zwanziger confinement scenario
\cite{Gribov:1978wm,Zwanziger:1994dh}.
In order to reduce the computational cost of
the simulations, we consider the pure $SU(2)$ gauge theory,
including studies in three (instead of
the usual four) space-time dimensions.

Besides the topics described below, we
also carry out numerical studies of: gauge-fixing algorithms
\cite{gfix}, Gribov-copy effects \cite{Cucchieri:1997dx},
the chiral phase transition of QCD with two dynamical fermions 
\cite{Tuca}, the equation of state
of spin models with Goldstone modes \cite{ON},
cluster percolation \cite{Wand_BJP} and short-time dynamics for spin
models \cite{OP_wand}.


\section{Confinement scenarios in Landau gauge}
\label{sec:conf}

As said above, the study of the infra-red (IR) limit of QCD is of central
importance for understanding the mechanism of
confinement and the dynamics of partons at low energy.
Despite being non-gauge-invariant,
gluon and ghost propagators are powerful tools in the
(non-perturbative) investigation of this limit
\cite{Alkofer:2000wg,Holl:2006ni}.
In fact, according to the Gribov-Zwanziger
\cite{Gribov:1978wm,Zwanziger:1993dh,
Sobreiro:2004us} and to the Kugo-Ojima \cite{Kugo}
confinement scenarios in Landau gauge,
the ghost propagator must show a divergent behavior
in the IR limit --- stronger than
$p^{-2}$ --- for vanishing momentum $p$.
This strong IR divergence corresponds to a long-range interaction in
real space, which may be related to quark confinement.
At the same time, according to the former scenario, the gluon propagator
must be suppressed and may go to zero in the IR limit
\cite{Gribov:1978wm,Zwanziger:1993dh,
Sobreiro:2004us,Zwanziger}.
This would imply that the real-space gluon propagator
is positive and negative in equal measure, i.e.\ reflection positivity is maximally
violated \cite{Cucchieri:2004mf,Furui:2004cx}. As a consequence,
the Euclidean 2-point function cannot represent the correlations of
a physical particle. This result may be viewed as an indication of
gluon confinement \cite{Alkofer:2000wg}.

These theoretical predictions have been confirmed by studies using
Dyson-Schwinger equations
(DSE's) \cite{Alkofer:2000wg,Holl:2006ni}.
In particular, studies of DSE's in Landau gauge have found
\cite{Zwanziger:2001kw,Lerche:2002ep,Alkofer:2003jj}
an IR behavior of the form $G(p) \sim p^{-2 \kappa -2} = p^{-2 a_G -2}$
for the ghost propagator and of the form $D(p) \sim
p^{4 \kappa - 2} = p^{2 a_D - 2} $ for the gluon propagator
with the same exponent $\kappa$ (i.e.\ with $a_G = a_D / 2$).
In 4d one usually finds $\kappa \gtrsim 0.5$ for pure $SU(N_c)$ gauge theory.
Note that $\kappa > 0.5$ implies $D(0) = 0$.
For the $3d$ case the exponents are $a_G \approx 0.4$ and $a_D \approx 1.3$.
Note that in the $d$ dimensional case
\cite{Zwanziger:2001kw,Lerche:2002ep} the relation between $a_D$ and $a_G$
is given by $a_D = 2 a_G + (4 - d)/2$, implying for the quantity
$\alpha_s(p) = (g^2 / 4 \pi) D(p) G^2(p) p^6$ the IR behavior
$p^{2 (a_D - 2 a_G)} = p^{4 - d}$. Thus, in the 4-dimensional case the
running coupling $\alpha_s(p)$ displays and IR fixed point.

Numerical studies of lattice gauge theories confirm the IR divergence
of the Landau ghost propagator
\cite{Suman,Cucchieri:1997dx,Bloch:2003sk} 
and an IR suppression of the gluon propagator.
More precisely, a decreasing gluon propagator at small momenta
has been obtained for the $3d$ $SU(2)$ Landau case using very large lattices
\cite{Cucchieri:2003di}
and --- recently --- in the $4d$ $SU(3)$ Landau case with the use of
asymmetric lattices \cite{Oliveira}.
Similar results has also been obtained for the the equal-time three-dimensional transverse
gluon propagator in $4d$ $SU(2)$ Coulomb gauge
\cite{Coulomb}.
In this last case, one also obtains an excellent fit of the
transverse propagator by a Gribov-like formula.
Finally, direct support to the Gribov-Zwanziger
and to the Kugo-Ojima scenarios has been presented in \cite{Cucchieri:1997ns}
and in \cite{Furui}, respectively.

Thus, the two nonperturbative approaches above seem to support the
Gribov-Zwanziger and the Kugo-Ojima confinement scenarios in
Landau gauge. However, the agreement between the two methods is
still at the qualitative level.
Moreover, recent lattice studies \cite{Boucaud:2005ce,Ilgenfritz:2006gp}
seem to indicate a null IR limit for $\alpha_s(p)$,
instead of a finite nonzero value.
At the same time, a study based on 
DSE's \cite{Fischer:2005ui} showed that torus and continuum
solutions are qualitatively different. This suggests a
nontrivial relation between studies on compact and on noncompact manifolds
and could have important implications for lattice studies.
Also, the nonrenormalizability of the ghost-gluon vertex
--- proven at the perturbative level \cite{Taylor:1971ff},
confirmed on the lattice \cite{Mihara,Ilgenfritz:2006gp}
(for $p \gtrsim 200$ MeV) and used in
DSE studies to simplify the coupled set of equations
--- has been 
recently criticized in Ref.\ \cite{Boucaud:2005ce}. Thus, 
clear quantitative understanding of the two confinement scenarios is
still an open problem.


\section{Infinite-volume limit}

The study of the IR behavior of propagators and vertices,
i.e.\ for momenta smaller than 1 GeV,
requires careful consideration of the infinite-volume limit.
Indeed, since the smallest non-zero momentum that
can be considered on a lattice is given by $p_{min} \approx 2 \pi /L$
--- where $L$ is the size of the lattice in physical units ---
it is clear
that one needs to simulate at very large lattice sizes in order to probe the
small-momentum limit.
The consideration of very large lattice sizes
requires parallelization and high efficiency of the code
in order to obtain good statistics in the Monte Carlo
simulation. Thus, an optimized parallel code is of great
importance \cite{Cucchieri:2003zx}.
Our numerical code is parallelized using MPI;
for the random number generator we use a double-precision
implementation of RANLUX (version 2.1) with luxury level set to 2.  


\subsection{Very large lattice side}
\label{sec:large}

In Ref.\ \cite{Cucchieri:2003di} we have evaluated the lattice
gluon propagator $D(k)$ and study it as a function of the
momentum $ p(k) $ in the 3d $SU(2)$ Landau case, using data from
the largest lattice side to date, i.e.\ up to $140^3$.
This allowed us to consider momenta as small as 51 MeV
(in the deep IR region) and physical lattice sides almost
as large as 25 fm.

\begin{figure}[t]
\begin{center}
\vspace*{-4cm}
\epsfxsize=0.40\textwidth
\leavevmode\epsffile{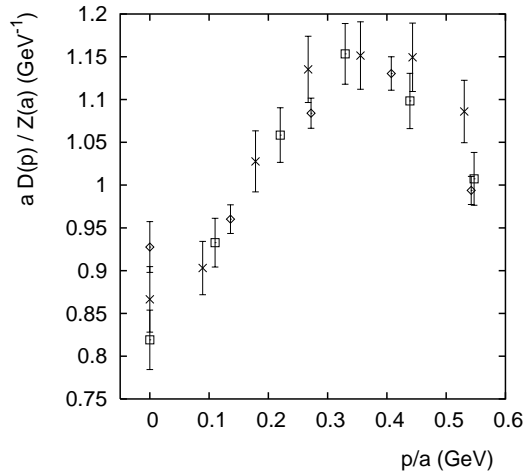}
\protect\vskip -0.5cm
\end{center}
\caption{~Plot of the rescaled gluon propagator $a D(0) / Z(a)$ as
a function of the momentum for
$V=80^3$ and $\beta=4.2\,(\times)$, $\,5.0\,(\Box)$, $6.0\,(\Diamond)$.
Error bars are obtained from propagation of errors.
}
\label{fig:fits}
\end{figure}

In order to compare lattice data at different $\beta$'s, we
apply the matching technique described in 
\cite{Leinweber:1998uu}, i.e.\ the propagators are multiplied by
a factor $Z(a)$ depending on $\beta$ or, equivalently, on the
lattice spacing $a$.
The method works very well (see Fig.\ \ref{fig:fits});
indeed, data obtained using different $\beta$ values nicely
collapse into a single curve.
We find that the gluon propagator decreases in
the IR limit for momenta smaller than $p_{max}$, which
corresponds to the mass scale $M$ in a Gribov-like
propagator $D(p) = p^2 / (p^4 + M^4)$.
From the plot we can estimate
$p_{max} = M = 0.8^{+0.2}_{-0.1} \sqrt{\sigma} = 350^{+100}_{-50} $ MeV,
in agreement with Ref.\ \cite{Cucchieri:1999sz}.
(Here $\sigma$ is the string tension.)

\begin{figure}[b]
\begin{center}
\vspace*{-3.6cm}
\epsfxsize=0.40\textwidth
\leavevmode\epsffile{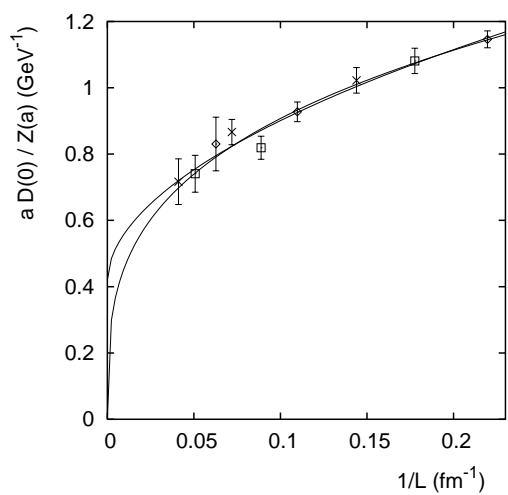}
\protect\vskip -0.5cm
\end{center}
\caption{~Plot of the rescaled gluon propagator
at zero momentum as a function of the inverse lattice side.
We also show the fit of the data using the Ansatz
$d + b / L^c$, both with $d = 0$ and with $d \neq 0$.
Error bars are obtained from propagation of errors.
}
\label{fig:zero}
\end{figure}

In Fig.\ \ref{fig:zero} we plot the rescaled gluon
propagator at zero momentum, namely $a D(0) / Z(a)$, as a
function of the inverse lattice side $L^{-1} = 1/(a N)$ in physical
units (fm$^{-1}$). We see that $a D(0) / Z(a)$
decreases monotonically as $L$ increases, in agreement with
Ref.\ \cite{Bonnet:2001uh}. It is interesting
to notice that these data can be well fitted using the
simple Ansatz $d + b / L^c$ both with $d=0$ and $d \neq 0$
(see Figure \ref{fig:zero}). In order to decide for one or the
other result one should go to significantly larger lattice sizes.

Also, in Ref.\ \cite{Cucchieri:2003di} we have shown that
the data for the gluon propagator are well fitted by Gribov-like
formulae, yielding an IR critical exponent $\kappa \approx 0.65$
in agreement with recent analytic results (see Section \ref{sec:conf}).
Recently \cite{inprep} we have extended this analysis to the ghost propagator,
considering lattice volumes up to $80^3$ for the
coupling $\beta = 4.2$.
A fit to the data using the fitting function $b/p^a$
(in the interval $p \leq 0.5$ GeV)
gives $a = 2.40(2)$.
This result would imply $a_G \approx 0.4$,
also in agreement with the results reported in Section \ref{sec:conf} above.


\subsection{Asymmetric lattices}

Recently, very asymmetric lattices \cite{Oliveira,
Parappilly:2006si} have been considered
in order to explore the IR limit od QCD. As a test of this method,
we have extended \cite{Cucchieri:2006za}
the gluon propagator study presented in \cite{Cucchieri:2003di}
[for the $3d$ $SU(2)$ case in minimal Landau gauge],
by including results for the ghost propagators from very large
lattices. At the same time, we evaluated the propagators using also asymmetric lattices,
in order to verify possible systematic effects related to the use
of asymmetric lattices (as suggested in \cite{Ilgenfritz:2006gp}), 
by comparing the results to the ones obtained for symmetric 
lattices.

\begin{figure}[t]
\begin{center}
\includegraphics[height=0.75\hsize]{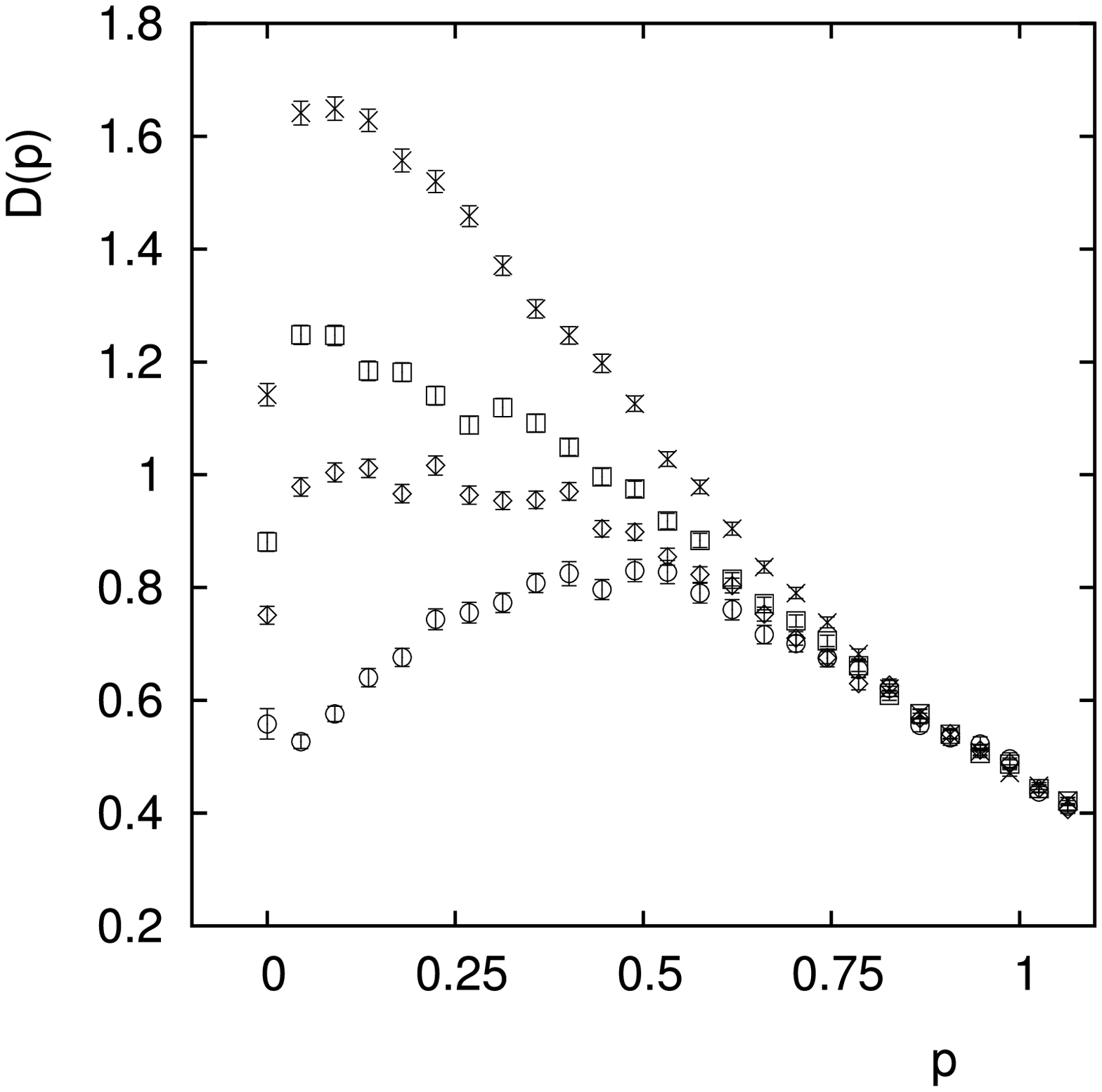}
\protect\hskip -1.2cm
\includegraphics[height=0.75\hsize]{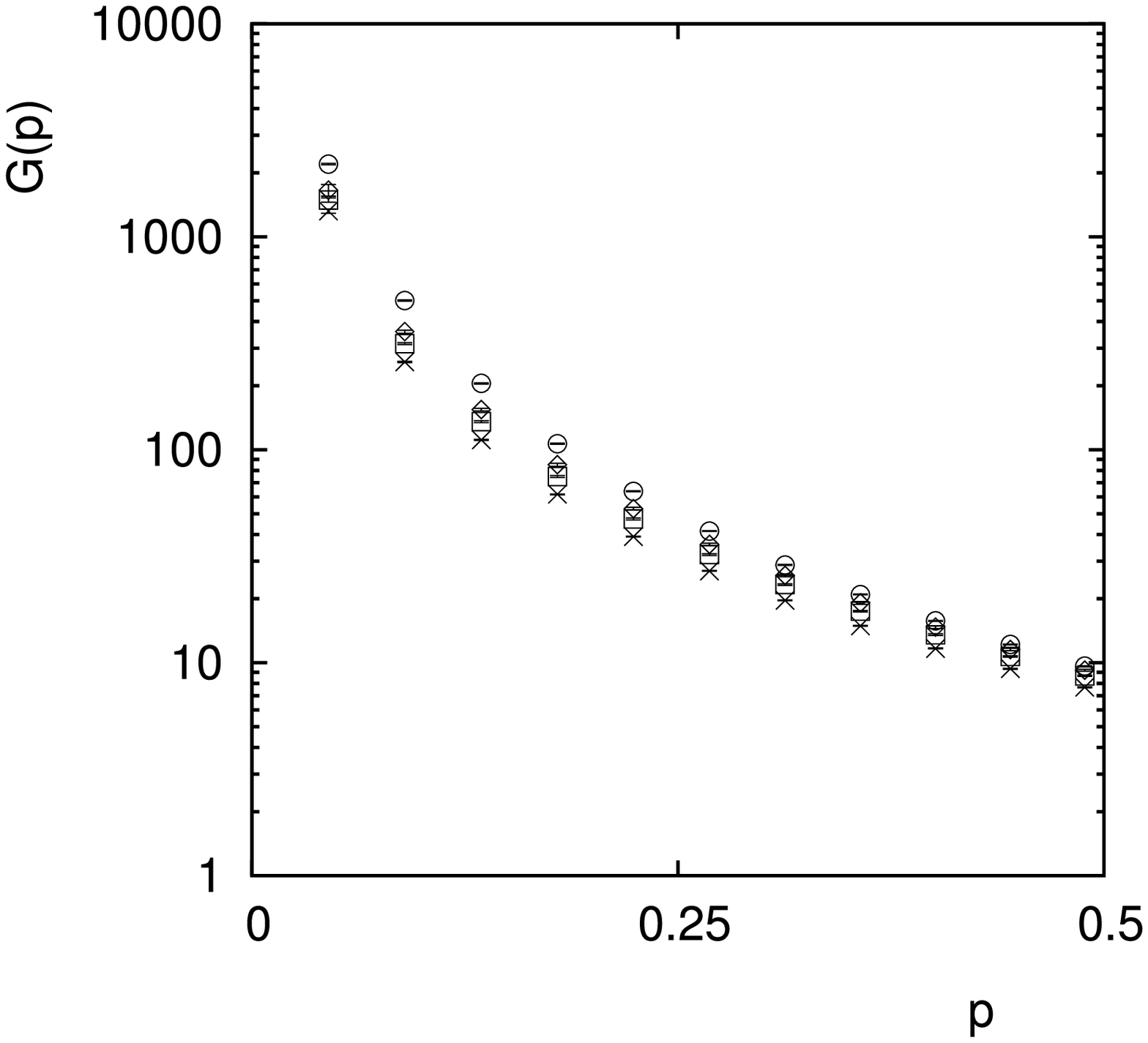}
\protect\vskip -1.1cm
\end{center}
\caption{Plot of the gluon propagator $D(p)$ (top)
and of the ghost propagator $G(p)$ (bottom) as a function of $p$
for lattice volumes $V = 8^2 \times 140\,(\times)$, $12^2 \times 140
\, (\Box)$, $16^2 \times 140 \, (\Diamond)$ and $V = 140^3 \, (\bigcirc)$.
All quantities are in lattice units.
Note the logarithmic scale on the $y$ axis in the bottom plot.
Errors represent one standard deviation.
\label{fig:gluonghost}
}
\end{figure}

We find, for both propagators, clear evidences of systematic effects
at relatively small momenta, i.e.\ $p \lesssim 1.5 \sqrt{\sigma}
\approx 650 \mbox{MeV}$. 
In particular, the gluon (respectively, ghost)
propagator is less suppressed (respec.\ enhanced) in the IR limit
when considering asymmetric lattices than for the case of symmetric lattices
(see Fig.\ \ref{fig:gluonghost}).
This implies that the estimates for the IR critical exponents $a_G$
and $ a_D $ are systematically smaller in the asymmetric case compared
to the symmetric one.

Also, for the gluon propagator and considering the asymmetric lattices,
one would estimate
a value $M \lesssim 0.25/a \approx 0.22 $ GeV
as a turnover point in the IR, i.e.\ the momentum $p_{max} = M$
for which the propagator reaches its peak.
On the other hand, considering the
largest (symmetric) lattice volumes,
i.e.\ $ V = 140^3 $
(see the top plot in Fig.\ \ref{fig:gluonghost}), the gluon propagator is clearly a decreasing
function of $p$ for $p \lesssim 0.5/a$, corresponding to
$M \lesssim 0.435$ GeV. This is in agreement with the result
reported in Section \ref{sec:large} above.
We thus see a difference of almost a factor 2 between the
momentum-turnover point in the symmetric and asymmetric cases.

Finally, we have seen that the extrapolation to infinite volume of results 
obtained using asymmetric lattices is also most likely affected by systematic
errors.
We conclude that, even though using an asymmetric lattice does not modify the
qualitative behavior of the two propagators, one should be careful
in extracting quantitative information from such studies.

\section{Reflection-positivity violation}

The relation between reflection positivity and Euclidean 
correlation functions can be made explicit by considering
the spectral representation \cite{Alkofer:2000wg,Aiso:au}
\begin{equation}
D(p)\;=\; \int_0^{\infty}\,dm^2\,\frac{\rho(m^2)}{p^2 + m^2}
\end{equation}
for the Euclidean propagator in momentum space.
Then, the statement of reflection positivity is equivalent to a 
positive spectral density $\rho(m^2)$.
This implies that the temporal correlator at zero spatial momentum
$D(t, {\bf p}=0)$ can be written as
\begin{equation}
C(t)\;\equiv\;D(t,0)\;=\; 
\int_0^{\infty}\,d\omega\,\rho(\omega^2)\,e^{-\omega\,t}\;.
\label{eq:Cdecomposition}
\end{equation}
Clearly, a positive density $\rho(\omega^2)$ implies that
$\,C(t)\;>\;0\,$.
Notice that having $C(t)$ positive for all $t$ does not ensure the positivity
of $\rho(\omega^2)$. On the other hand,
finding $C(t)<0$ for some $t$ implies that
$\rho(\omega^2)$ cannot be positive, suggesting 
confinement for the corresponding particle.

On the lattice, the real-space propagator can be evaluated
using
\begin{equation}
C(t) =  \frac{1}{N} \sum_{k_0=0}^{N-1}
e^{- 2\pi\,i k_0 t /N} \,D(k_0, 0) \;,
\label{eq:ltc}
\end{equation}
where $N$ is the number of points per lattice side and $D(k)$ is 
the propagator in momentum space. If the lattice action satisfies
reflection positivity \cite{Montvay:cy},
then we can write the spectral representation
\beqn
C(t) = \sum_n r_n e^{-E_n t} \;,
\label{eq:lattice_coft}
\eeqn
where $r_n$ are positive-definite constants.
Clearly, this implies that $C(t)$ is non-negative for all
values of $t$. 

Numerical indications of a negative real-space lattice Landau gluon
propagator have been presented in the $3d$ $SU(2)$ case
\cite{finiteTSU2},
in the magnetic sector of the $4d$ $SU(2)$ case at finite temperature
\cite{Cucchieri:2001tw} and, recently, in the $4d$ $SU(3)$ case
for one ``exceptional'' configuration \cite{Furui:2004cx}.

\begin{figure}[t]
\begin{center}
\includegraphics[height=0.75\hsize]{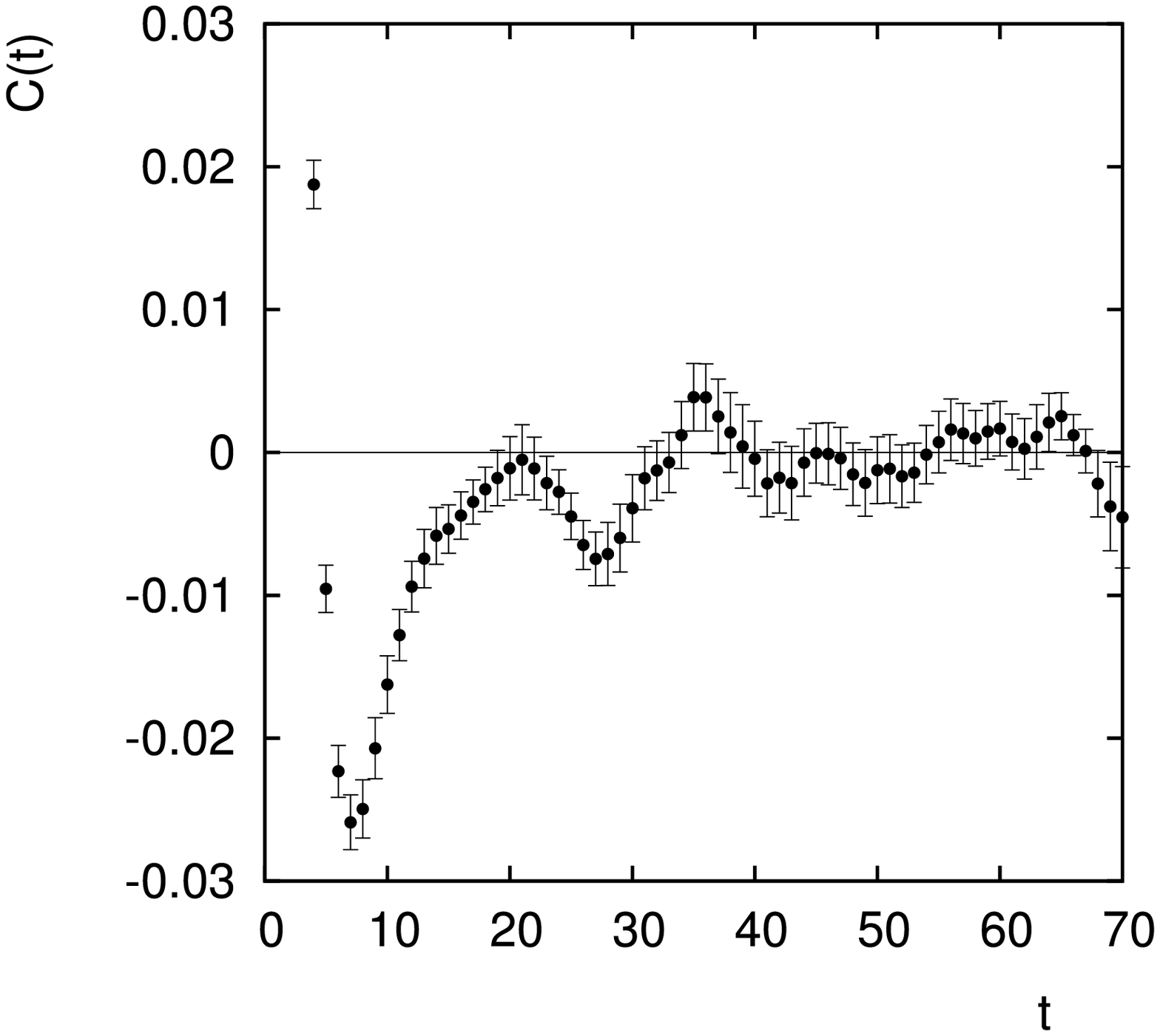}
\protect\vskip -0.4cm
\end{center}
\caption{Real-space propagator $C(t)$ as a function of $t$
for coupling $\beta=5.0$ and lattice volume $V = 140^3$.
Errors have been evaluated using bootstrap with 1000 samples.
All quantities are in lattice units.
\label{fig:redect_zoom}
}
\end{figure}

\begin{figure}[t]
\begin{center}
\includegraphics[height=0.75\hsize]{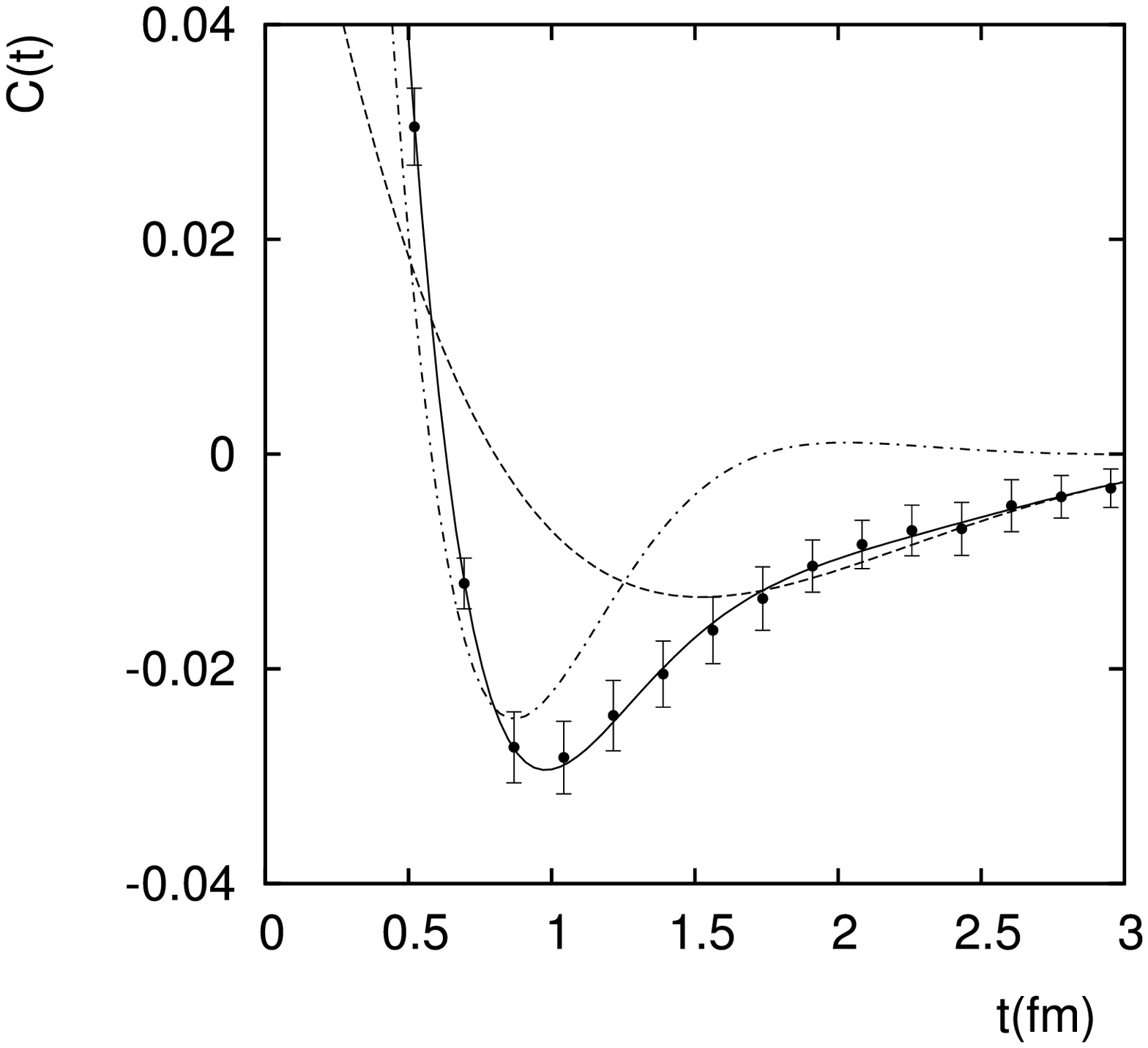}
\protect\vskip -0.4cm
\end{center}
\caption{Fit of ${C}(t)$ as a function of $t$ (in $fm$)
using a sum of two Gribov-like propagators
for lattice volume $V=140^3$ and coupling $\beta=4.2$. We also
display the two Gribov-like propagators separately.}
\label{fig:fitting}
\end{figure}

Using data from the largest lattice sides to date, we
verify explicitly (in the $3d$ case) \cite{Cucchieri:2004mf}
the violation of 
reflection positivity for the $SU(2)$ lattice Landau 
gluon propagator (see Fig.\ \ref{fig:redect_zoom}).
In particular, the propagator
becomes negative at $t \approx 0.7\,fm$ and the minimum
is reached at $t_{min} \approx 1\,fm$ (see Fig.\ \ref{fig:fitting}).
Note that the Gribov-like propagator
\begin{equation}
C(t) \,=\, \frac{e^{-M t / \sqrt{2}}}{2\,M}\,
     \cos\left(\frac{M t}{\sqrt{2}} \,
                      +\,\frac{\pi}{4}\right)
\label{eq:gribovreal}
\end{equation}
has its minimum at
$t_{min} = \pi / (M \,\sqrt{2}\,)$.
Thus, the above result for $t_{min}$ would imply
$M \approx \pi / \sqrt{2}\,fm^{-1} \approx 2.22\,fm^{-1} = 438\,MeV
= 0.995 \sqrt{\sigma} $, where $\sigma$
is the string tension.
Let us also recall that the momentum-space Gribov-like propagator
has its maximum at $p_{max} = M$ (see Section \ref{sec:large} above).
Moreover, we find that finite-size effects seem to become important only
at $t \gtrsim 3\,fm$. This means that our data for $t \in [0, 3]\,fm\,$ 
are essentially infinite-volume continuum results.
In the scaling region, the data are well described by 
a sum of Gribov-like formulas (see Fig.\ \ref{fig:fitting}),
with a light-mass scale 
$m\approx 0.74 \sqrt{\sigma} = 325\,MeV$.

It has been suggested \cite{Aubin:2003ih,Aubin:2004av} 
that the violation of spectral positivity in lattice Landau gauge be
related to the quenched auxiliary fields used for gauge fixing.
We note that the fitting form proposed for $C(t)$ in \cite{Aubin:2004av}
describes reasonably well our data up to $t=3\,fm$ --- yielding a
light-mass scale of about $1.14 \sqrt{\sigma} = 500\,MeV$ --- but cannot
account for the oscillatory behavior observed at very large
separations.


\section{Ghost-gluon vertex}

In the framework of quantum field theory, Faddeev-Popov 
ghosts are introduced in order to quantize non-Abelian gauge
theories. Although the ghosts
are a mathematical artifact and are absent from the physical
spectrum, one can use the ghost-gluon vertex and the ghost
propagator to calculate physical observables, such as the
QCD running coupling $\alpha_s(p)$, using the relation
\be
\alpha_s(p) \; = \; \alpha_0\,
  \frac{Z_3(p) \, \widetilde{Z}_3^2(p)}{\Z1^2(p)} \; .
\label{eq:alpha_run}
\ee
Here $\alpha_0 = g_0^2 / 4 \pi $ is the bare coupling constant and
$Z_3(p)$, $\widetilde{Z}_3(p)$ and $\Z1(p)$ are, respectively,
the gluon, ghost and ghost-gluon vertex renormalization functions.
The above
formula gets simplified if one considers the Landau gauge. Indeed,
in this case the vertex renormalization function $\Z1(p)$
is finite and constant, i.e.\ independent of the renormalization scale $ p $,
at least to all orders of perturbation theory
\cite{Taylor:1971ff,Piguet:1995er}.
Of course, it is important to verify in a non-perturbative way
that this result really holds. If it does, one can consider
in the Landau gauge
a definition of the running coupling constant that requires only
the calculation of the gluon and ghost propagators
\cite{vonSmekal}.

\begin{figure}[t]
\includegraphics[width=0.42\textwidth]{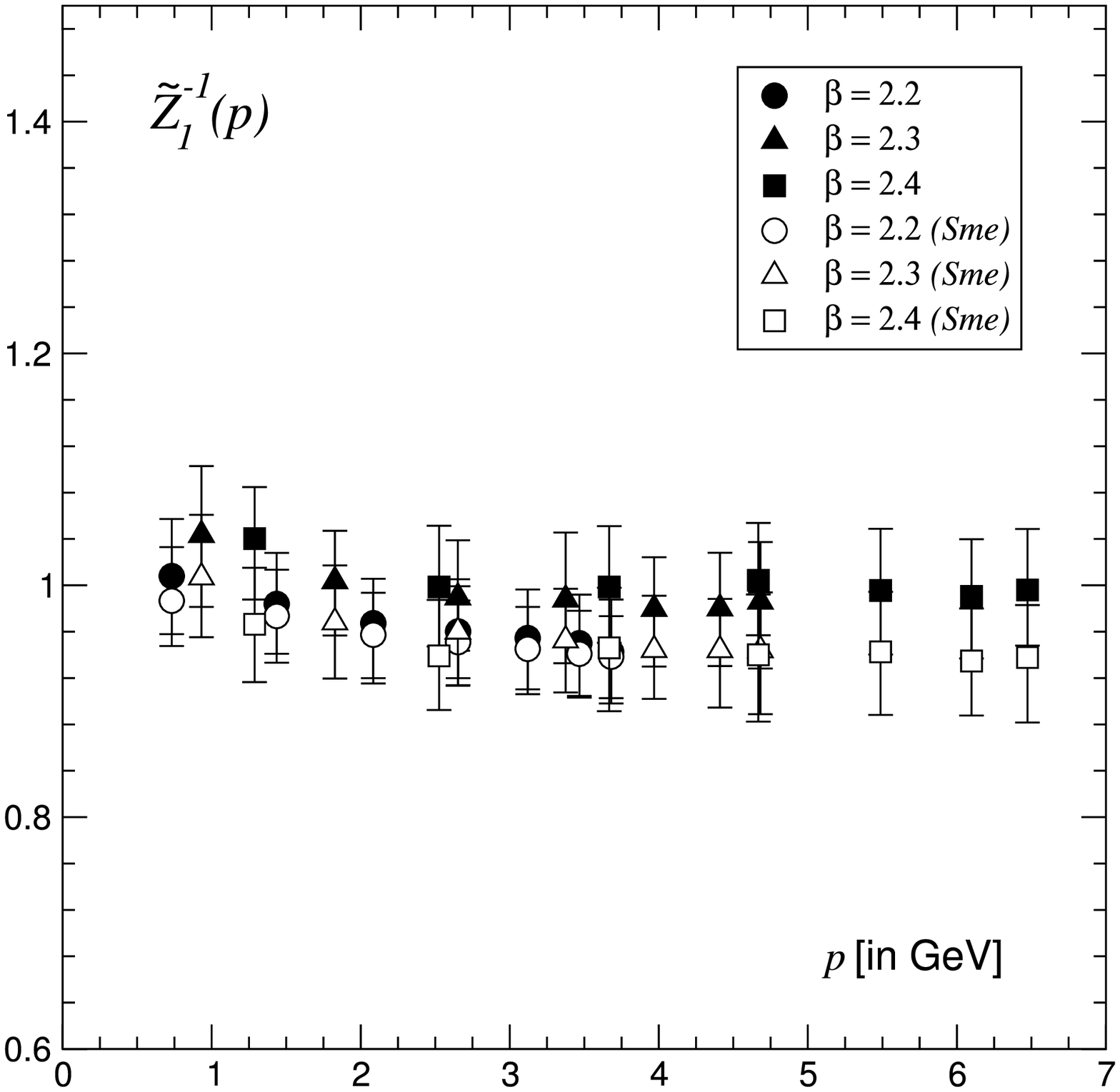}
\caption{Results for $\,\Z1^{-1}(p)\,$
for the lattice volume $V = 16^4$ as a function of $p = \hat{p}/a$ in GeV,
considering symmetric momenta, i.e.\ with 4 equal components.
We show data obtained using
two different gauge-fixing methods (with and without
the so-called smearing method \protect\cite{Hetrick:1997yy}).
Error bars were evaluated using the bootstrap method with 250 samples.}
\label{fig:Z}
\end{figure}

In Refs.\ \cite{Mihara}
we have studied the reduced ghost-gluon vertex function
$\Gamma^{abc}_{\!\mu}(0,p) $ and the
renormalization function $\Z1(p)$ in minimal Landau gauge
at the asymmetric point $(0; p, -p)$
in the $SU(2)$ case.
We find that the vertex function has the same
momentum dependence of the (lattice) tree-level vertex
--- i.e.\ $\sim \hat{p} \, \cos(\pi \, \widetilde{p} \, a / L) \sim
 \sin( 2 \, \pi \, \widetilde{p} \, a / L) $ --- and that
$\Z1(p)$ is approximately constant and equal
to 1, at least for momenta $p \gtrsim 1$ GeV
(see Fig.\ \ref{fig:Z}). This is 
a direct non-perturbative verification of the
perturbative result reported above.
In particular, using the result obtained at the
largest value of $\beta$ considered (i.e.\ $\beta = 2.4$)
we obtained $\Z1^{-1}(p) = 1.02^{+6}_{-7}$,
where errors include Gribov-copy effects and discretization
errors related to the breaking of rotational invariance.

\begin{figure}[t]
\begin{center}
\includegraphics[height=0.75\hsize]{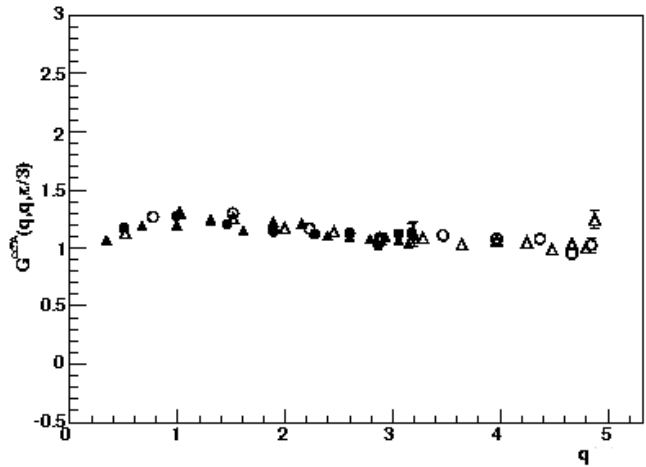}
\protect\vskip -0.4cm
\end{center}
\caption{The scalar quantity $G^{c\bar cA}(q,k,\phi)$, defined in
Eq.\ (27) of Ref.\ \cite{Cucchieri:2006tf},
as a function of the magnitude of the incoming anti-ghost momentum $q$.
Full symbols correspond to $\beta=4.2$ and open symbols to $\beta=6.0$;
circles are used for $V=20^3$ and triangles for $V=30^3$.
We plot data for the case with the three momenta equal.
}
\label{fig:ggv}
\end{figure}

Recently, this study has been extended (in the 3d case)
\cite{Cucchieri:2006tf} to other kinematical
configurations, including in particular the symmetric
point $p^2 = q^2 = k^2$. We find that the vertex
is essentially constant and of order one, for all momentum configurations
(see Fig.\ \ref{fig:ggv}),
confirming the results obtained in the 4d case
\cite{Mihara,Ilgenfritz:2006gp}.
This result is also in agreement with predictions from functional
methods in the 3d case
\cite{Schleifenbaum}.
In the same reference we also present the first numerical study of the
Landau-gauge three-gluon vertex in the 3d case and
results for the spectrum of the Faddeev-Popov operator, which plays
an important role in the Gribov-Zwanziger   
scenario \cite{Gribov:1978wm,Zwanziger:1993dh,Sternbeck:2005vs}.
In particular we have shown that (see Fig.\ 7 in Ref.\ \cite{Cucchieri:2006tf})
the smallest non-zero eigenvalue
of the Faddeev-Popov matrix goes to zero when the infinite-volume
limit is approached. As a consequence, in the continuum limit,
the average lattice
Landau configuration should belong to the first Gribov horizon, supporting the
Gribov-Zwanziger mechanism of confinement (see also \cite{Cucchieri:1997ns}).

\section{Ghost condensates}
\label{sec:ghostcond}

The QCD vacuum is known to be highly non-trivial at low energies \cite{Shuryak}.
This non-trivial structure manifests itself through the
appearance of vacuum condensates, i.e.\ vacuum expectation values of certain
local operators. In perturbation theory these condensates vanish,
but in the SVZ-sum-rule approach \cite{Shifman}
they are included as a parametrization of non-perturbative effects
in the evaluation of phenomenological quantities.
The two main such operators are $\,\alpha_s F_{\mu\nu} F^{\mu\nu} \,$ and
$\,m_q \overline{\psi}_q \psi_q$; their vacuum expectation values are the so-called
gluon and quark condensates. Both of these operators have mass dimension four.

In recent years, (gauge-dependent) condensates of mass dimension two have also received
considerable attention \cite{Gubarev:2000eu,Dudal:2002ye,
Lemes:2002ey,Dudal:2002xe,
Dudal:2003dp,Dudal,Sobreiro:2004us,
Dudal:2004rx,Dudal:2005bk,Capri:2005vw,
Kondo:2000ey,Kondo,Kondo:2001tm,Kondo:2003uq}.
In particular, the gauge condensate $\langle A_{\mu}^b A_{\mu}^b \rangle$
has been largely studied, since it should be sensitive to topological 
structures such as monopoles \cite{Gubarev:2000eu} and it
could play an important role in the quark-confinement scenario
through monopole condensation \cite{monopole}.
Moreover, the existence of a gauge condensate would imply a dynamical
mass generation for the gluon and ghost fields
\cite{Sobreiro:2004us,Dudal}.
Possible effects of the gauge condensate $\langle A_{\mu}^b A_{\mu}^b \rangle$
on propagators and vertices (in Landau gauge) have been studied through
lattice simulations in Refs.\
\cite{Boucaud,Boucaud:2005ce},
yielding $\langle A_{\mu}^b A_{\mu}^b \rangle \approx 3$ GeV$^2$.

\begin{figure}[b]
\begin{center}
\protect\hskip -1.2cm
\protect\vskip 0.4cm
\includegraphics[height=0.80\hsize]{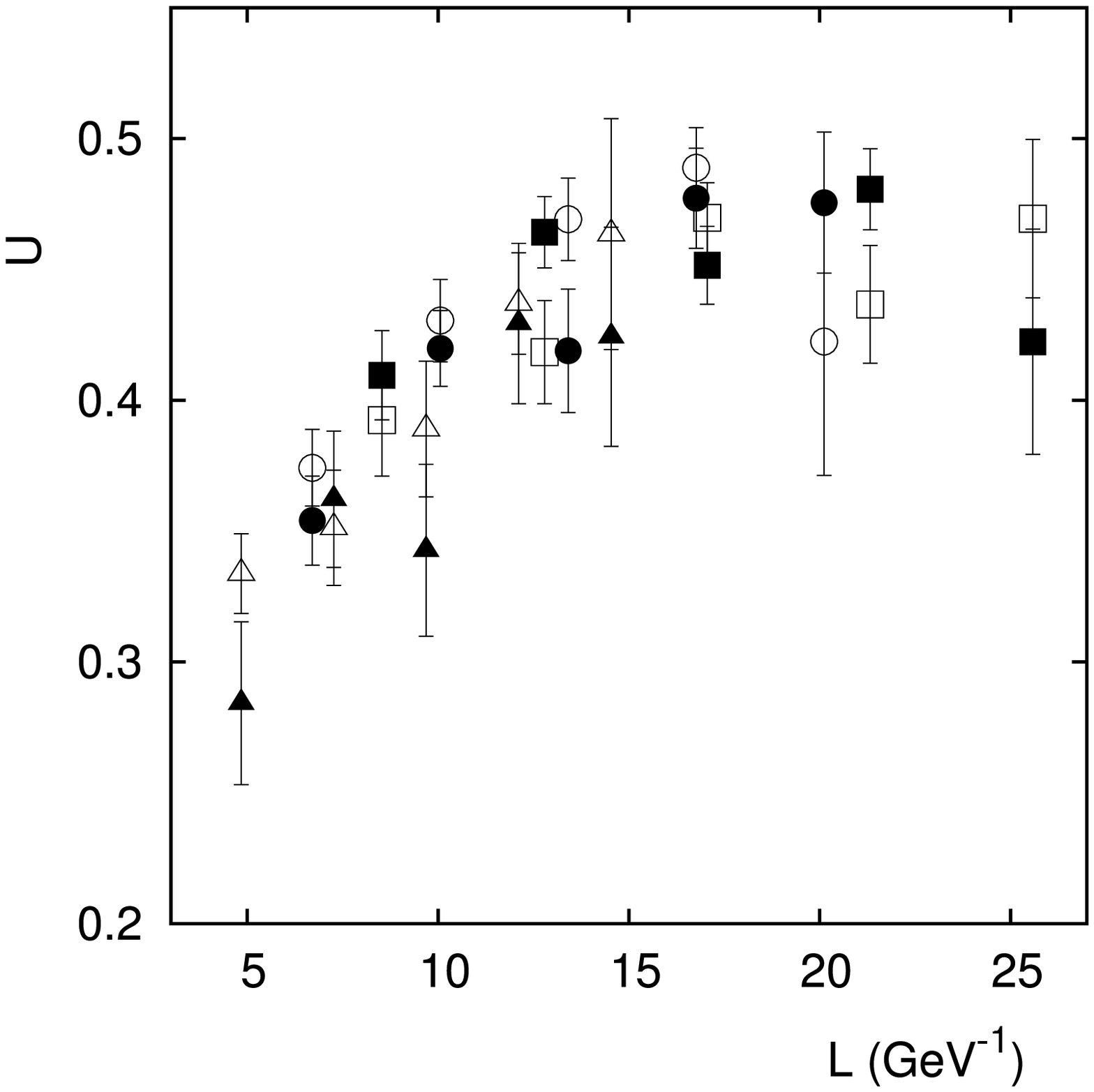} 
\caption{Results for the Binder cumulant $U$
for the quantity $\phi^b(p)$ as a function of the lattice
side $L$ (in GeV$^{-1}$) for various lattice volumes $V$ and
momentum with $\widetilde{p} = N/4$.
We show the data corresponding to asymmetric momenta [for $\beta = 2.2$
($\Box$), $\beta = 2.3$ ($\bigcirc$) and $\beta = 2.4$ ($\bigtriangleup$)]
and to symmetric momenta (with the corresponding filled symbols
for each $\beta$).
Errors have been estimated using the bootstrap method with 10,000 samples.
\label{fig:binder}
}
\end{center}
\end{figure}

Other vacuum condensates of mass dimension two considered by several groups
are the ghost condensates. These
condensates were first introduced in $SU(2)$ gauge theory in maximally
Abelian gauge (MAG) \cite{Schaden,Kondo:2000ey,Dudal:2002xe,Kondo:2003uq}.
More recently, the same condensates have been studied in other gauges
\cite{Kondo:2001tm,Dudal:2002ye}, such as 
the Curci-Ferrari and the Landau gauges.
In all cases it was found that the ghost condensates are related to
the breakdown of a global $SL(2,R)$ symmetry \cite{Alkofer:2000wg,Dudal:2002ye}.
In MAG the diagonal and off-diagonal components of the ghost propagators
are modified \cite{Lemes:2002ey,Kondo:2000ey} by ghost condensation.
Similar results have been obtained in other gauges \cite{Dudal:2003dp,Kondo:2001tm}.
In particular, in Landau gauge it was shown \cite{Dudal:2003dp}
that the off-diagonal (anti-symmetric) components of the ghost propagator $G^{cd}(p)$
are proportional to the ghost condensate $v$.

Finally, a mixed gluon-ghost condensate of mass dimension two has also been studied
by several authors \cite{Kondo,Kondo:2001tm,Dudal:2002xe,
Dudal:2004rx,Dudal:2005bk}, using various gauges. This mixed condensate is of
particular interest when considering interpolating gauges \cite{Dudal:2004rx}.
Indeed, it allows to generalize and relate results obtained in different gauges
for the gauge condensate $\langle A_{\mu}^b A_{\mu}^b \rangle$ and the ghost condensates.
Moreover, in MAG this mixed condensate would
induce a dynamic mass for the off-diagonal gluons \cite{Dudal:2005bk},
giving support to the Abelian-dominance scenario \cite{abelian}.
Thus,
the various gauge and ghost condensates could all play an
important role in the dual superconducting scenario of
quark confinement \cite{supercond},
being related to monopole condensation and to Abelian dominance.

In Ref.\ \cite{Cucchieri:2005yr} we carried out a thorough investigation of ghost condensation in the
so-called Overhauser channel for pure $SU(2)$ Yang-Mills theory in minimal Landau gauge.
In particular, we evaluate numerically the off-diagonal components of the ghost
propagator $G^{cd}(p)$ as a function of the momentum $p$. We find that
$ \langle \phi^b(p) \rangle = \epsilon^{bcd}\, \langle G^{cd}(p) \rangle / 2 $
is zero within error bars, but with large fluctuations.
At the same time,
we see clear signs of spontaneous breaking of a global symmetry,
using the quantity $\phi^b(p)$ as an order parameter.
As in the case of continuous-spin models in the
ordered phase (see for example \cite{MC2}),
spontaneous symmetry breaking is supported by
two (related) observations: 1) by comparing the statistical 
fluctuations for the quantities $\phi^b (p)$ and $|\phi^b (p)|$;
2) from the non-Gaussian shape of the 
statistical distribution of $\phi^b (p)$, which can be observed
by considering a histogram of the data or by evaluating the
so-called Binder cumulant (see Fig.\ \ref{fig:binder}).
Since, in Landau gauge, the vacuum expectation value of the
quantity $\phi^b (p)$ should be proportional (in the Overhauser channel) to the ghost
condensate $v$ \cite{Dudal:2003dp}, it seems reasonable
to conclude that the broken symmetry is the $SL(2,R)$ symmetry,
which is related to ghost condensation \cite{Dudal:2002ye,Alkofer:2000wg}.
This interpretation has been recently criticized in Ref.\ 
\cite{Furui:2006rx}.
There, the nonzero value obtained for the Binder cumulant has been explained
by considering multi-dimensional Gaussian distributions and a modified
definition for the Binder cumulant. We note however that
the order parameter considered in our study, i.e.\
the magnitude of $\phi^b (p)$, is a scalar quantity. Therefore,
in analogy with studies of $O(N)$-vector models (see for example
\cite{O4}), the standard
definition of the Binder cumulant should apply to our study as well.

Let us note that, from our data, the Binder cumulant $U$ seems
to be approximately null at small lattice volume and
to converge to a value $U \approx 0.45$ for physical lattice side $L \gtrsim 15$
Gev$^{-1} \approx 3$ fm, corresponding to a mass scale of less than $100$ MeV.

In Ref.\ \cite{Cucchieri:2005yr} we have also shown that
the sign of $\phi^b(p)$ is
related to the sign of the Fourier-transformed gluon field components
${\widetilde A}(q)$ and that
$\phi^b (p)$ has discretization effects similar to
those obtained for the ghost-gluon vertex \cite{Mihara}.
Then, using the rescaled quantity
$|\, L^2 \, \phi^b (p) / \cos(\pi\tilde{p} a / L) \,|$,
we find (at small momenta) a behavior $p^{-z}$ with
$z\approx 4$, in agreement with analytic predictions \cite{Dudal:2003dp}.
On the other hand, from our fits we find
that the ghost condensate $v$ is consistent with zero within error bars, i.e.\
the quantity $|\,L^2 \phi^b (p) / \cos(\pi\tilde{p} a / L) \,|$ does not
approach a finite limit at small momenta, at least for $p \geq 0.245\,$ GeV.
Using the Ansatz
\beq
G^{cd}(p) \;=\; \frac{p^2 \, \delta^{cd}\, + \, v\,\epsilon^{cd}}
{p^4 + v^2} \,\mbox{,}
\label{fpm0}
\eeq
we obtain for the ghost condensate the upper bound $v \ll 0.058$ GeV$^2$.
More precisely, our data rule out values of $v$ greater than
$0.058$ GeV$^2 \approx (240 $ MeV$)^2$ but would still be
consistent with a ghost condensate
$v \lesssim 0.025$ GeV$^2 \approx (160 $ MeV$)^2$.
Let us note that, in analytic studies
\cite{Dudal:2002xe,Dudal:2003dp,Capri:2005vw,Sawayanagi:2003dc} one finds that
the ghost condensate induces a tachyonic gluon mass proportional to
$\sqrt{v}$, which modifies the dynamic gluon mass related
to the gauge condensate $\langle A_{\mu}^b A_{\mu}^b \rangle$. 
Thus, one should expect a relatively small ghost condensate
$\,v\,$ in order to obtain a global (non-tachyonic) gluon mass.
Let us recall that a dynamic gluon mass of the order of a few
hundred MeV has been considered in several phenomenological
studies \cite{mass}.
A similar mass scale was also obtained in
numerical studies of the gluon propagator in Landau gauge
\cite{Cucchieri:1999sz,Cucchieri:2003di,Cucchieri:2004mf}.


\section{Running coupling}
\label{sec:runn}

\begin{figure}[t]
\vspace*{-4.4cm}
\begin{center}
\epsfxsize=0.43\textwidth
\leavevmode\epsffile{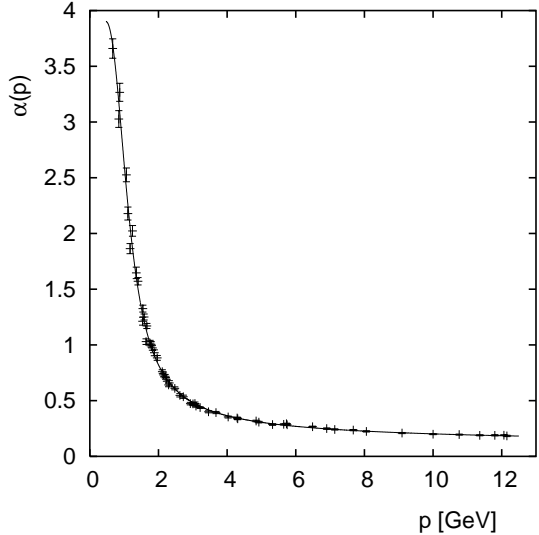}
\caption{Fit for the running coupling $\alpha_s(p)$
using eq.\ (\ref{eq:alphaSC}) with
$C = 0.072(8)$,
$a = 1.9(3)$, $\Lambda = 1.31(1)$
and $m = 1.0(6)$.
}
\label{fig:alphaSC}
\end{center}
\end{figure}

Of great importance for phenomenological purposes is the running coupling
strength $\alpha_s(p)$ \cite{vonSmekal} defined in Eq.\
(\ref{eq:alpha_run}) above. In particular,
this quantity enters directly the quark Dyson-Schwinger equation
(DSE) and can be interpreted
as an effective interaction strength between quarks \cite{Bloch:2002eq}.
Let us note that, working in Landau gauge and
in the momentum-subtraction scheme, the
running coupling (\ref{eq:alpha_run}) can be written as
\cite{vonSmekal}
\be
\alpha_s(p) \; = \; \alpha_0\, F(p)\,J^2(p) \; ,
\label{eq:alpha_run2}
\ee
where $F(p)$ and $J(p)$ are, respectively,
the gluon and the ghost form factors and we used
the result $\Z1(p) \,=\, 1$.
As explained in Sec.\ \ref{sec:conf} above, studies using DSE's
have found that, if the IR sum rule $2 a_G - a_D = 0$ is satisfied,
then this running coupling develops a fixed point in the IR limit
(see for example \cite{Lerche:2002ep})
\be
\lim_{p\to 0} \alpha_s(p) \; = \;
\alpha_c \;=\;\hbox{constant},
\label{eq:alphac}
\en
with $\alpha_c \approx 8.92/N_c$
in the SU($N_c$) case for $\kappa \approx 0.596$.

This quantity has been studied numerically by several groups (see for example
\cite{Furui:2004cx,running,Boucaud:2005ce}).
In Ref.\ \cite{Bloch:2002we}
we have evaluated this running coupling constant and tried a fit to the data
using the fitting function
\begin{equation}
\alpha_s(p) \,=\, C \, p^4 / \left[ (p^4 + m) \, s (a) \right]
\label{eq:alphaSC} 
\end{equation}
where $s(a) = (11 / 24 \pi^2) \log{[1 + (p^2/\Lambda^2)^{a}]}$.
Note that if $a = 2$ one finds a fixed point in the IR limit
equal to $\alpha_c = 24 \pi^2 C \Lambda^4 / 11 m$.
Also, this fitting functions satisfies the  
leading ultraviolet behavior $\sim 1 / \log{[(p^2/\Lambda^2)]}$
for the running coupling. Results are reported in Fig.\ \ref{fig:alphaSC};
we find $a = 1.9(3)$ and $\alpha_c = 24 \pi^2 C \Lambda^2 / 11 m
\approx 4.6$, in agreement with the result reported above.
Similar results have also been obtained in \cite{Bloch:2003sk}.
We stress that in Refs.\ \cite{Bloch:2002we,Bloch:2003sk}
we have compared results obtained using two slightly different
lattice formulations, yielding consistent results in all cases considered.

\begin{figure}[t]
\begin{center}
\includegraphics[height=0.92\hsize]{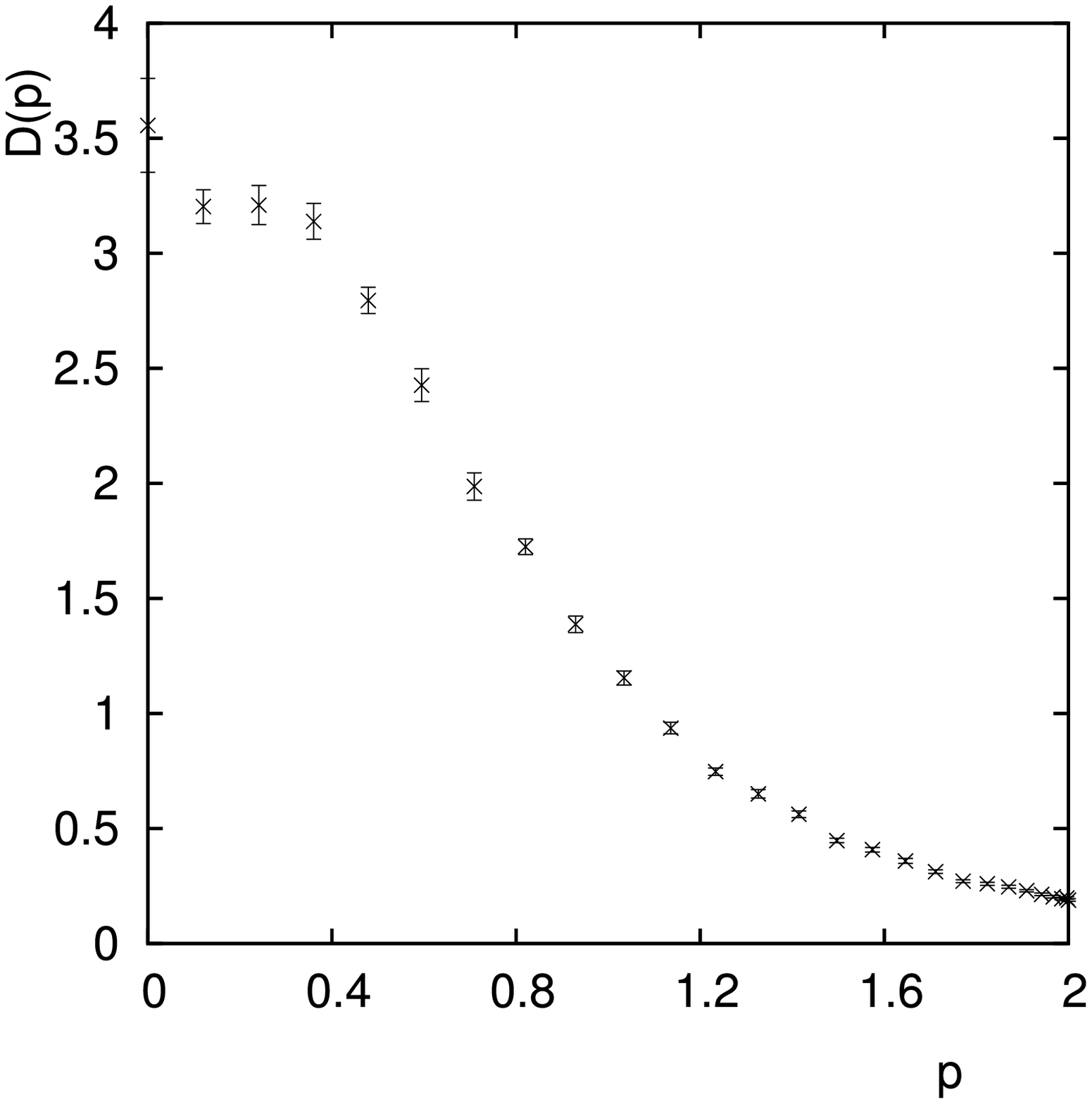}
\protect\vskip -1.1cm
\end{center}
\caption{Plot of the gluon propagator $D(p)$
as a function of $p$ for lattice volume
$V = 52^4$ at $\beta = 2.2$.
All quantities are in lattice units (here $a \approx 1$ GeV$^{-1}$).
Errors represent one standard deviation.
The larger value at $p=0$ is probably due to an insufficient number of
gauge-fixing iterations.
\label{fig:largeV}
}
\end{figure}

As said in Section \ref{sec:conf} above,
recent lattice studies \cite{Boucaud:2005ce,Ilgenfritz:2006gp}
seem to indicate a null IR limit for $\alpha_s(p)$,
instead of a finite nonzero value $\alpha_c$. We should stress, however,
that in these studies special care must be taken in order
to eliminate finite-size effects, especially when the IR region
is considered. Indeed, we know that, if the
the gluon propagator is not suppressed at small momenta,
the true infrared regime is surely not reached yet.
This is a difficult numerical task since we have checked that in the
4d $SU(2)$ case a lattice volume $V = 52^4$ at $\beta = 2.2$,
corresponding to a physical lattice side of about 10 fm, is still
not sufficient to show a decreasing propagator in the limit of
small momenta (see Fig.\ \ref{fig:largeV}).
We note that in Fig.\ 1 of \cite{Boucaud:2005ce} and
in Fig.\ 3 of \cite{Ilgenfritz:2006gp} the running coupling 
starts to decrease for a momentum $p \approx 0.5$ GeV.
Thus, from our discussion in Section \ref{sec:large} above, considering
the results shown in Fig.\ \ref{fig:largeV} and also the results
reported in \cite{Leinweber:1998uu,Bonnet:2001uh}, we can say that
the data considered in \cite{Boucaud:2005ce,Ilgenfritz:2006gp}
for momenta $p \lesssim 500$ MeV are most likely affected by strong
finite-size effects.


\section{Conclusions}

The Gribov-Zwanziger
and the Kugo-Ojima confinement scenarios in Landau gauge
are well supported, at the qualitative level, by several
studies based on different analytical and numerical approaches.
On the other hand, as we have shown here, a comprehensive
analysis of these confinement scenarios at the quantitative level
using lattice numerical simulations could represent a very challenging
task.


\section*{ACKNOWLEDGMENTS}

The authors thank the organizers for the invitation.
Research supported by 
Funda\c{c}\~ao de Amparo \`a Pesquisa do Estado de S\~ao Paulo (FAPESP)
(under grant \# 00/05047-5).
Partial support from Conselho Nacional de Desenvolvimento
Cient\'{\i}fico e Tecnol\'ogico (CNPq) is also acknowledged.


\end{document}